# A phase-locked laser system based on modulation technique for atom interferometry

Wei Li, Xiong Pan[*], Ningfang Song, Xiaobin Xu, Xiangxiang Lu

*Institute of Opto-electronics and Technology, School of Instrumentation Science and Opto-electronics Engineering, Beihang University, Beijing, 100191, China*

[*]e-mail: 08768@buaa.edu.cn

**Abstract** We demonstrate a Raman laser system based on phase modulation technology and phase feedback control. The two laser beams with frequency difference of 6.835 GHz are modulated using electro-optic and acousto-optic modulators, respectively. Parasitic frequency components produced by the electro-optic modulator are filtered using a Fabry-Perot Etalon. A straightforward phase feedback system restrains the phase noise induced by environmental perturbations. The phase noise of the laser system stays below -125 $rad^2$/Hz at frequency offset higher than 500 kHz. Overall phase noise of the laser system is evaluated by calculating the contribution of the phase noise to the sensitivity limit of a gravimeter. The results reveal that the sensitivity limited by the phase noise of our laser system is lower than that of a state-of-art optical phase-lock loop scheme when a gravimeter operates at short pulse duration, which makes the laser system a promising option for our future application of atom interferometer.

## 1 Introduction

Atom interferometers have been widely used for precision metrology, such as measurement of the Newtonian gravitational constant [1, 2], the fine-structure constant [3], and testing the equivalence principle [4]. Their high sensitivity to inertial forces ensures that atom interferometers can be used as accelerometers [5] and gravimeters [6]. Raman atom interferometers usually use two laser beams that induce coherent Raman transitions between two hyperfine levels of the alkali atoms to manipulate the atomic wave packets [7]. Phase difference between the two beams, referred to as beat note phase in this paper, is incorporated in the phase of an atom interferometer. Therefore, laser beams with ultra-low phase noise associated with beat note are essential to ensure high performance of an atom interferometer. Typically, there are two methods to generate the laser beams. One method, referred to as optical phase-lock loop (OPLL), utilizes two lasers with beat note phase-locked to an ultra-stable reference oscillator [8--11]. The intensity ratio and polarization of the laser beams can be controlled independently. However, an intricate phase servo system is essential in order to achieve low residual phase noise and wide locking bandwidth of the OPLL. Complex analog + digital phase and frequency detector is designed to ensure that the phase detector has both the broad capture range of digital circuits and the high speed and low noise of analog mixers [12]. Two or three control paths are employed in order to provide high gain in low frequency as well as high bandwidth which is limited by long cables and the response time of the laser diode current controller [9, 13]. The other method based on phase modulation technology uses a high frequency acousto-optic modulator (AOM) or electro-optic modulator (EOM) to generate the required frequency difference between the two laser beams [7, 14--16]. In earlier schemes, the two phase modulated laser beams are separated in different optical paths, resulting in severe degeneration of phase noise due to vibration of optical paths. Recently, O. Carraz et al. [17] demonstrate a compact laser system at 780 nm by doubling the frequency of a telecom fiber bench at 1560 nm. The Raman lasers, generated by adding sidebands on the slave laser with a fiber EOM, are following the same optical path and thus phase noise that induced by both the frequency noise of the laser and vibrations of optical paths is avoided. Besides, the power ratio of the two laser beams (0th and +1th) can be controlled in order to avoid AC stark shift. However, the unwanted sideband produces an additional couple of Raman laser, giving systematic error that can limit the sensitivity of an atom interferometer. A model used to precisely calculate the phase shift by the additional laser lines is presented and validated by experiment [18].

In this article, we present a laser system based on both phase modulation and phase feedback control. Firstly, we describe the methods of phase modulation used in the two laser paths. The parasitic frequency components induced by phase modulation are measured and suppressed in order to avoid phase shift in an atom interferometer. Then, we introduce the phase feedback system (PFS) used to restrain the phase noise caused by environmental perturbations. The advantages of our PFS compared with OPLL scheme are discussed. Finally, we validate the laser system by calculating the sensitivity limit of a gravimeter due to phase noise of the laser beams. We also propose some methods in order to further improve the performance of our laser system.

## 2 Laser system

Our set-up consists of two laser beams having a frequency offset of 6.835 GHz, that will be used to induce stimulated Raman transitions from the ground state $|F = 1, m_F = 0\rangle$ to the state $|F = 2, m_F = 0\rangle$ of $^{87}$Rb in the future. The laser system is illustrated in Fig. 1. An external cavity diode laser (ECDL, Toptica, DL pro) is used for our laser source whose typical linewidth and mode-hop free tuning range are 100 kHz and 50 GHz, respectively.

Small part of the optical power of the ECDL is sent to a saturated absorption spectrometer (SAS). The frequency of the ECDL is locked to the D2 line of $^{87}$Rb ($F_g = 2$ to $F_e = 3$) and a red frequency detuning of 1.3 GHz is achieved by double-passing the laser beam through a AOM driven by microwave signal of 650 MHz (not shown in the figure for simplicity). The other optical power is amplified from 30 mW to 1.3 W by a tapered amplifier (TA), and then sent to a modified Mach–Zehnder interferometer. The split and recombination of the laser beams is achieved using two polarizing beam splitters (PBS). Half waveplates are used before the two PBS in order to regulate the power distribution ratio of the two laser beams.

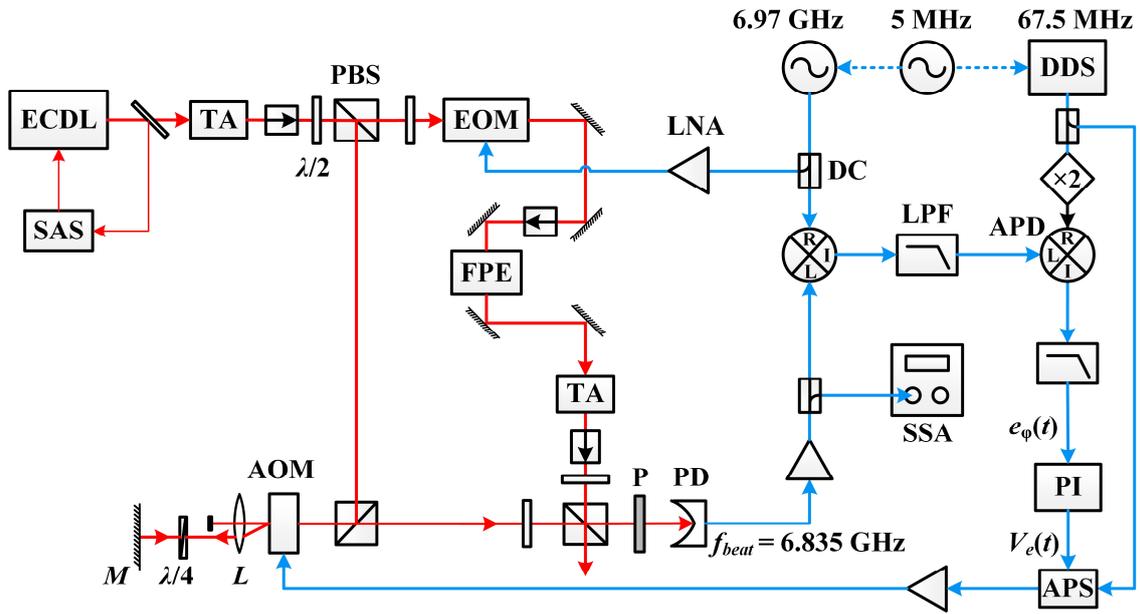

Fig. 1 Scheme of the experimental system set-up. The red and blue line represent laser beams and electronic signals, respectively. λ/2: half waveplate. λ/4: quarter waveplate. M: reflection mirror. P: polarizer. PD: photodetector. SSA: signal source analyzer. LPF: low pass filter. APD: analog phase detector. DC: directional coupler. LNA: low noise amplifier. DDS: direct digital synthesizer. APS: analog phase shifter
.

### 2.1 Phase modulated by EOM

The upper laser with power of 600 mW is phase modulated by a resonant EOM (Newport, 4851) driven by microwave signal of 6.97 GHz. The EOM consists of an electro-optic crystal in a resonant microwave cavity that ensures efficient phase modulation and low drive-voltage requirement. In order to avoid amplitude modulation, a half waveplate is used to orientate the polarization of the laser beam along the electro-optic axis of the crystal. Imposing a sinusoidal phase modulation at frequency $\Omega$ and a peak phase modulation depth $m$ on the EOM, the phase variation is $\phi(t) = m\sin(\Omega t)$. The resulting electric field of the optical beam can be written as:

$$E = E_0 e^{i(\omega t + m\sin(\Omega t))}$$
$$= E_0 \sum_{k=-\infty}^{\infty} J_k(m) e^{i(\omega + k\Omega)t} \quad (1)$$

The fraction of optical power transferred into the $k$th sideband is proportional to $[J_k(m)]^2$, where $J_k(m)$ is the $k$th-order Bessel function and $J_{-k}(m) = (-1)^k J_k(m)$. The output beam is the combination of the carrier and sidebands that are spaced at multiples of the modulation frequency about the carrier. Limited by the threshold input power of the EOM, a microwave (6.97 GHz) with power of 35 dBm is applied, corresponding to a phase modulation depth of 0.72. The calculated percentages of optical power transferred into the needed $+1^{th}$ sideband and remaining in the carrier ($0^{th}$) are 11.25% and 76.7%, respectively. In order to get the pure $+1^{th}$ sideband, a temperature controlled Fabry-Perot Etalon (FPE) is placed after the EOM as an optical filter. The FPE is made of fused silica with thickness of 20 mm and both sides coated with a reflection of 96%. The reflection mirrors around the FPE make is easier to adjust the orientation of the laser beam and obtain maximum transmittance. The optical isolator before the FPE absorbs the reflected optical power. Compared with Fabry-Perot cavity constructed with two confocal mirrors, the FPE has higher transmittance and lower sensitivity to vibrations. Transmission spectrum of the FPE is obtained by linearly scanning the optical frequency with a triangular wave voltage applied on piezoelectric transducer (PZT) of the ECDL. The spectrum is shown in Fig. 2.

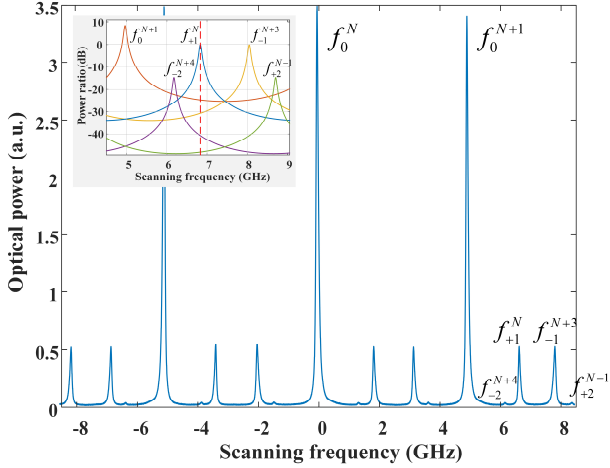

**Fig. 2** Transmission spectrum of the FPE. The superscript and subscript represent the interference order of the FPE and order of sideband of the EOM, respectively. The inset shows the power ratios of carrier and sidebands to $+1^{th}$ sideband.

The abscissa is transformed from scanning time into scanning frequency according to the interval between the carrier and $+1^{th}$ sideband which corresponds to the modulation frequency of 6.97 GHz. A free spectral range of 4.96 GHz is precisely measured. The modulation efficiency represented by percentages of optical power transferred into the major sidebands is shown in Table. 1. Only 11.1% of the optical power is transferred into the desired $+1^{th}$ sideband and 77.2% of then remains in the carrier. The experimental data well corresponds to the calculated results. Then we calculate the transmittances of each of the frequency components when the transmission frequency of FPE is set at the $+1^{th}$ sideband. The results are listed in the second column of Table. 1. The last column of the table shows the final power ratios of carrier and major sidebands to $+1^{th}$ sideband after the FPE.

Table. 1 Modulation efficiencies, transmittances and Power ratios of carrier and different sidebands

|         | Modulation efficiencies (%) | Transmittances (dB) | Power ratios (dB) |
|---------|-----------------------------|---------------------|-------------------|
| $f_{+2}$ | 0.4 | -33 | -47.4 |
| $f_{+1}$ | 11.1 | 0 | 0 |
| $f_0$ | 77.2 | -33 | -24.6 |
| $f_{-1}$ | 10.9 | -31 | -31.1 |
| $f_{-2}$ | 0.38 | -26 | -40.7 |

It can be found that the parasitic frequency components mostly come from the transmitting carrier, which is approximately 25 dB less than $+1^{th}$ sideband. The results are much smaller than that in Ref. [17] where no optical filter is used. In addition, the modulation frequency of 6.97 GHz is 135 MHz detuning from the transition frequency of 6.835 GHz, which further reduces the impact of carrier to the phase shift in an atom interferometer. Considering the insertion losses of the EOM (1.8 dB), optical isolator (1 dB) and the FPE (3 dB), we obtain an optical power of 15 mW of the $+1^{th}$ sideband transmitting from the FPE. Then a second TA amplifies the laser power to approximately 400 mW, that is enough for our future applications

**2.2 Phase modulated by AOM**

The nether laser with initial optical power of 700 mW is sent to an AOM (AA OPTO-ELECTRONIC) with center frequency at 70 MHz and bandwidth of 30 MHz. The AOM is placed in cat's eye configuration in order to reduce the sensitivity of the alignment to the value of the frequency. A radio frequency signal of 67.5 MHz drives the AOM to produce an optical frequency offset of 135 MHz of the laser beam. An optimum diffraction efficiency of 80% is achieved when the AOM is driven by electrical power of 32 dBm and a double-passing efficiency of 60% is finally obtained. Frequency chirp, used to compensate the Doppler frequency shift induced by gravity in our future application for atom interferometry, can be achieved by scanning the frequency of the driving signal of the AOM. The optical power ratio between the two laser beams can be controlled by altering electrical power of the driving signal. Moreover, as will be described in Sect. 3, the phase error signal from the PFS is feedback into the laser beam by modifying the phase of the driving signal, achieving a closed phase-locked loop. Then, the two intersecting linearly polarized laser beams with a frequency difference of 6.835 GHz are recombined in another PBS. Most of the optical power passes through the beam splitter, and approximately 250 μW of power from each laser beam is coupled into a photodetector (PD, New Focus, 1544-B) with a bandwidth of 12 GHz. The beat note of the two laser beams is transformed into electrical signal with frequency of 6.835 GHz and then used for measurement of phase noise and phase feedback.

Our optical scheme, where the laser beams are phase modulated separately in two optical paths, provides flexible control of optical power ratio and frequency offset between the two laser beams. However, there is a trade-off between the flexibility and the phase noise performance of the laser system. When the two lasers beams are separated in different optical paths, frequency noise of the ECDL will be converted into the phase noise of beat note if the two optical paths have different propagation distances [19]. In addition, the phase noise of beat note would also be degenerated seriously by environmental perturbations: vibration of optical devices, air current and temperature variation etc. In our previous work [20], the difference between propagation distances have been removed by using a technique named Frequency Modulated Continual Wave (FMCW) [21], meaning that the frequency noise of the ECDL becomes negligible compared with the other phase noise sources.

## 3 Phase feedback system

In order to decrease the adverse effect due to environmental perturbations, a phase feedback system (PFS) is build, as illustrated in the right part of Fig. 1. The reference signal of 6.97 GHz and the signal of 67.5 MHz from the direct digital synthesizer (DDS) are both phase-locked to a 5 MHz ultra-high performance crystal oscillator (Microsemi, 1000C), whose phase noise spectral density at frequency offset of 1 Hz from the carrier reaches a value of -127 rad$^2$/Hz. Part of electric power of the two reference signals is amplified and used to drive the EOM and AOM. The remaining power is used as reference signal for extracting the phase error induced by environmental perturbations of the optical paths. The beat note of 6.835 GHz is amplified and then frequency mixed with the reference signal of 6.97 GHz using a two-tone-terminator mixer (Marki, T3-20). Small part of power of the beat note (10dBm) is sent to a signal source analyzer for phase noise measurement. Then, after the low pass filter, an analog phase detector (Mini-Circuits, ZX05) is used to determine the phase difference between the down-converted signal of 135 MHz and the reference signal of 135 MHz obtained by directly doubling the signal of 67.5 MHz from the DDS. The low pass filter filters out the redundant signals, and the remaining phase error signal is monitored by an oscilloscope (not shown in the figure). During the process of phase error extraction, the common noise originating from the reference sources is eliminated, meaning that PFS has no restraint on the phase noise of the reference sources. Therefore, the phase noise of the beat note mainly consists of two parts: the noise caused by environmental perturbations, which is restrained by the PFS, and the noise from the reference sources. In order to feed the phase error back into the optical system, most schemes use a slave laser [8--11] or a voltage controlled oscillator [7] as the final controlling element. The locking processes involve frequency-locking and phase-locking. In our system, however, the frequency difference between the laser beams is quite stable since it is strictly determined by the reference signals. Hence, directly feeding the phase error back into the phase of laser beam is a much more straightforward option. Here, the driving signal of the AOM is first phase shifted by an analog phase shifter (APS, SigaTek, SF28A7) and then amplified to 32 dBm using a low noise amplifier (Mini-Circuits, ZHL-03-5WF). The APS, with a maximum phase shift of 500 degrees, converts the voltage of the error signal into phase shift of the driving signal and finally shifts the phase of the nether laser beam. The error signal is amplified and integrated by a proportional-integral controller, and then used as the control voltage of the APS. In addition, we add a lag-lead compensator to counteract the inherent phase shift of the APS and improve the stability of the PFS. Compared with other systems where independent lasers are used, our phase feedback system needs only phase-locking of the loop which reduces the complexity of the feedback loop. Besides, the required bandwidth is much smaller since the main noise that our loop must suppress is acoustic noise caused by environmental perturbations.

## 4 Results and discussion

A signal source analyzer (SSA, Agilent E5052B) is used to measure the phase noise of the 6.835 GHz beat note signal. Since the intrinsic phase noise sensitivity of the SSA is insufficient to measure the beat note, we enhance it by taking advantage of the correlation function of the SSA (a correlation of N times can improve the sensitivity by 5log (N) dB). The results are shown in Fig. 3.

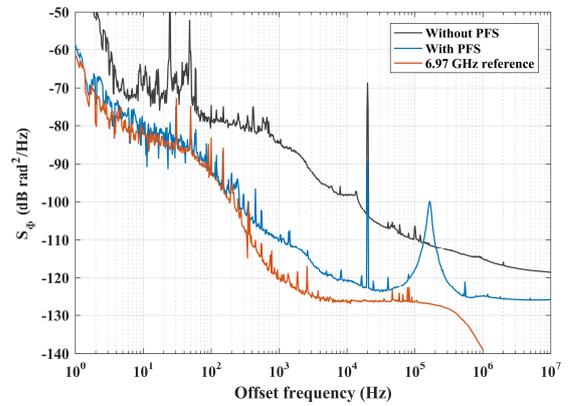

**Fig. 3** Phase noise spectral density of the beat note signal of 6.835 GHz. Black curve: without phase feedback. Blue curve: with phase feedback. The red curve is the phase noise of the reference signal of 6.97 GHz.

The red curve in Fig. 3 represents phase noise spectral density of the reference signal of 6.97 GHz. The small bump near 100 Hz is caused by phase-lock loop (PLL) of the 6.97 GHz frequency chain. Phase noise of the DDS is not shown in the figure because it is much smaller than the reference signal of 6.97 GHz at the whole frequency range. Phase noise spectral density of the 6.835 GHz beat note without and with phase feedback are shown in Fig. 3 by the black and blue curves, respectively. One can see that below 400 Hz, the phase noise spectral density of the beat note with phase feedback almost overlaps the phase noise spectral density of the reference signal of 6.97 GHz, indicating that the phase noise caused by environmental perturbations is strongly suppressed and can be neglected compared with the phase noise of the reference signal in this frequency range. Above 400 Hz, the loop gain of the PFS decreases rapidly, resulting in an insufficient suppression of the phase noise. Hence, the total phase noise in this frequency regime is dominated by noises caused by environmental perturbations. The sharp peak at 20 kHz can be explained by the frequency modulation of the saturated absorption frequency stabilization of the ECDL. The phase noise of the beat note is not affected by the frequency modulation since the difference between the two propagation distances has been removed, as described in Sec. 2. However, the FPE, acting as an optical frequency filter, converts the optical frequency modulation into intensity modulation, which in turn appears as a fake peak in the phase noise curve. This mechanism is confirmed by directly measuring the optical intensity of the laser beam transmitting from the FPE. FFT analysis of the optical intensity reveals that a sharp peak also appears at frequency of 20 kHz and the intensity fluctuation induced by frequency modulation reaches a value of 0.4 %. The bump closed to 200 kHz in Fig. 3 indicates a much lower bandwidth of the PFS compared with OPLL schemes. However, the range of this bump is only 300 kHz that is much smaller than several megahertz in OPLL schemes [9, 10]. The phase noise stays under -125 rad$^2$/Hz at frequency offset higher than 500 kHz. A comparison of the phase noise spectral density curves with and without PFS shows that the phase noise caused by environmental perturbations is highly suppressed by the PFS. Specifically, the rejection ratio reaches a value of more than 20 dB at frequencies lower than 10 kHz, and the overall phase noise spectral density at an offset of 30 kHz is approximately -123 rad$^2$/Hz.

Considering the future application for atom interferometer, we evaluate the phase noise performance of the laser system by calculating the root mean square (rms) phase noise $\Delta\Phi$ using a weight function that takes into account the pulse duration $\tau_R$ and the interferometric time $T$ of a $\pi/2 - \pi - \pi/2$ sequence. The weight function, as derived in [22] is expressed by

$$|H(\omega)|^2 = \left| -\frac{4\Omega_R \omega}{\omega^2 - \Omega_R^2} \sin\left(\frac{\omega(T+2\tau_R)}{2}\right) \right.$$
$$\left. \times \left[ \cos\left(\frac{\omega(T+2\tau_R)}{2}\right) + \frac{\Omega_R}{\omega}\sin\left(\frac{\omega T}{2}\right) \right] \right|^2 \quad (2)$$

where $\Omega_R = \pi/2\tau_R$ is the Rabi oscillation frequency. The weight function acts as a low-pass filter, with an effective cutoff frequency $f_0 = \sqrt{3}/12 \cdot \tau_R^{-1}$. The rms phase noise $\Delta\Phi$ is given by [22]

$$\Delta\Phi^2 = \int_0^{+\infty} |H(2\pi f)|^2 S_\Phi(f) df \quad (3)$$

where $S_\Phi(f)$ is the phase noise spectral density of the laser system. Similar with Ref. [10], we calculate the sensitivity limit of gravity measurement due to the overall phase noise of the laser system. The sensitivity limit $\Delta g/g$ is given by [10]

$$\frac{\Delta g}{g} = \frac{\Delta\Phi}{k_{eff}T^2 g} \quad (4)$$

where $k_{eff}$ is the effective wavenumber and $g$ is gravitational acceleration. Given a typical interferometric time $T = 150$ ms, the sensitivity limit varies with different pulse duration $\tau_R$. As shown in Fig 4, the blue curve represents the sensitivity limit due to the overall phase noise of the laser system with PFS, and the red curve reveals the sensitivity limited by the phase noise of our 6.97 GHz reference.

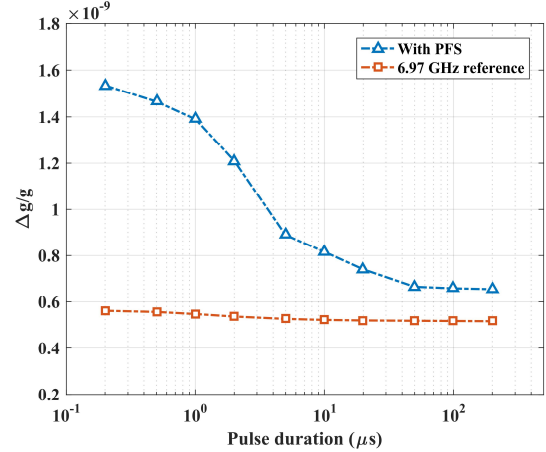

**Fig 4**. Sensitivity limits corresponding to different pulse duration $\tau_R$. The interferometric time $T$ equals 150 ms for all the data.

At long pulse duration $\tau_R = 100$ $\mu$s, the sensitivity limit is 3.4 times higher than that in [10], reaching a value of $6.6 \times 10^{-10}$. At this condition, the phase noise of the 6.97 GHz reference at low frequencies contributes mostly to the sensitivity limit, which is $5.2 \times 10^{-10}$. Considering that the cut off frequency of the weight function becomes higher as the pulse duration decreasing, the contribution of high frequency phase noise to the sensitivity limit turns out to be dominant at short pulse duration. For instance, the increment of sensitivity limit when pulse duration decreases from 100 $\mu$s to 1 $\mu$s is approximately $7.4 \times 10^{-10}$, which, however, is 3.5 times smaller than that in Ref. [10]. Due to the lower phase noise at high frequencies, the sensitivity limited by the phase noise of our laser system in turn becomes twice lower than that in Ref. [10] at short pulse duration of 1 $\mu$s.

There are several methods to enhance the phase noise performance of the system. Firstly, both reducing the number of free space optics and compacting the laser system will significantly reduce the sensitivity of phase noise to environmental perturbations. For example, lots of reflectors and spaces can be saved by integrating the EOM and FPE into a single mount. Secondly, restraining the bump near 100 Hz caused by the PLL of the 6.97 GHz frequency chain can improve the phase noise at low frequencies. Finally, the bandwidth of PFS, mainly determined by the phase-frequency characteristics of the electronic devices, should be increased. We use frequency response analyzer (NF, FRA5087) to test the frequency response of electronic components. The result reveals that the major component restricting the bandwidth of PFS is the APS, whose inherent phase lag increases rapidly after 50 kHz. The phase lag deteriorates the phase-frequency characteristics and decreases the bandwidth of the PFS. In future studies, the APS will be replaced by one that maintain small phase lag in the frequency range of several megahertz, which can significantly increase the bandwidth and reduce the high frequency phase noises.

Of course, there are other aspects of our laser system that affect the performance of an atom interferometer: intensity stability, wavefront quality, mode-overlap of the two laser beams and so on. The intensity fluctuation of the upper laser beam is approximately 1% in a period time of 1 hour, mainly caused by fluctuation of temperature and therefore change of transmission of the FPE. The intensity of the nether laser beam is much more stable because only one AOM is used in this path. Nevertheless, an intensity stabilization system is being built in order to active control the intensity of the two laser beams. The wavefront quality and mode-overlap can be improved by using mode filters or coupling the two laser beams into a single mode optical fiber. In addition, an AOM will be used to switch and generate the pulse sequence in our future application of atom interferometer.

## 5 Conclusion

To conclude, we have built a low phase noise Raman laser system that can be used in $^{87}$Rb atom interferometers. The laser system is the combination of a modified Mach-Zehnder interferometer and a phase feedback system. Phase noise induces by frequency noise of ECDL is removed by decreasing the difference between propagation distances of the two laser beams. The Mach-Zehnder interferometer, with EOM and AOM in each of the paths, provides flexible adjustments of intensity ratio and frequency scanning between the two laser beams. A temperature controlled FPE filters out the parasitic frequency components, resulting in a power ratio of more than 24 dB between the required sideband and the parasitic components. Phase noise induced by environmental perturbations is strongly restrained by the PFS. In contrast to OPLL schemes, we sidestep the frequency-locking process by directly feeding the phase error back into the optical system via an APS, which significantly decreases the complexity of the feedback loop.

The phase noise spectral density of the 6.835 GHz beat note at frequency offset of 30 kHz reaches a value of approximately -123 rad$^2$/Hz and stays below -125 rad$^2$/Hz at frequency offset higher than 500 KHz. We evaluate the global phase noise performance by calculating the sensitivity limit of a gravimeter with typical parameters. The results reveals that the sensitivity limit due to the phase noise of our laser system is twice lower than that of a state-of-art OPLL scheme when a gravimeter operates at short pulse duration of 1 $\mu$s, which makes our laser system an alternative choice for highly precise atom interferometers.

**Acknowledgments** The authors would like to thank Jixun Liu and Yixiang Yu for fruitful discussions and Shifeng Yang for his contribution to the construction of the laser system.